\documentclass[12pt]{article}
\RequirePackage{amsthm,amsmath,amsbsy,amsfonts}
\usepackage{bm,graphicx,psfrag,epsf}
\usepackage{enumerate}
\usepackage{natbib}
\usepackage{paralist, booktabs, multirow}
\usepackage{url}
\usepackage[colorlinks,citecolor=blue,urlcolor=blue]{hyperref} 
\newcommand{\blind}{0}

\newtheorem{proposition}{Proposition}[section]
\newtheorem{remark}{Remark}

\numberwithin{equation}{section}
\theoremstyle{plain}

\usepackage{amssymb}
\usepackage[usenames]{color}
\usepackage[mathscr]{eucal}
\usepackage{bm}
\usepackage{xspace}
\usepackage{paralist}
\usepackage{ifpdf}
\usepackage{graphicx}
\usepackage{subfigure}
\usepackage{tabulary}
\usepackage{threeparttable}
\usepackage[linesnumbered,boxed]{algorithm2e}
\usepackage{ifpdf}

\def\X{{\bf X}}
\def\Y{{\bf Y}}
\def\Z{{\bf Z}}
\def\K{{\bf K}}
\def\L{{\bf L}}
\def\H{{\bf H}}
\def\I{{\bf I}}
\def\be{{\bm \beta}}
\def\bSigma{{\bf \Sigma}}
\def\bGamma{{\bf \Gamma}}
\def\bPi{{\bf \Pi}}

\def\bI{{\mathbb I}  }
\def\1{{\bf 1}}
\def\0{{\bf 0}}

\def\cS{{\cal S}}

\def\cM{{\cal M}}
\def\cB{{\cal B}}
\def\cL{{\cal L}}

\def\cF{{\cal F}}
\def\cD{{\cal D}}

\def\bbR{\mathbb{R}}

\ifpdf
\newcommand{\Sec}[1]{\hyperref[sec:#1]{Section~\ref*{sec:#1}}} 
\newcommand{\App}[1]{\hyperref[sec:#1]{Appendix~\ref*{sec:#1}}} 
\newcommand{\Eqn}[1]{\hyperref[eq:#1]{{\rm (\ref*{eq:#1})}}} 
\newcommand{\Part}[1]{\hyperref[part:#1]{(\ref*{part:#1})}} 
\newcommand{\Fig}[1]{\hyperref[fig:#1]{Figure~\ref*{fig:#1}}} 
\newcommand{\Tab}[1]{\hyperref[tab:#1]{Table~\ref*{tab:#1}}} 
\newcommand{\Thm}[1]{\hyperref[thm:#1]{Theorem~\ref*{thm:#1}}} 
\newcommand{\Lem}[1]{\hyperref[lem:#1]{Lemma~\ref*{lem:#1}}} 
\newcommand{\Prop}[1]{\hyperref[prop:#1]{Proposition~\ref*{prop:#1}}} 
\newcommand{\Cor}[1]{\hyperref[cor:#1]{Corollary~\ref*{cor:#1}}} 
\newcommand{\Def}[1]{\hyperref[def:#1]{Definition~\ref*{def:#1}}} 
\newcommand{\Alg}[1]{\hyperref[alg:#1]{Algorithm~\ref*{alg:#1}}} 
\newcommand{\Ex}[1]{\hyperref[ex:#1]{Example~\ref*{ex:#1}}} 
\newcommand{\As}[1]{\hyperref[as:#1]{Assumption~{\rm\ref*{as:#1}}}} 
\newcommand{\Reg}[1]{\hyperref[as:#1]{Condition~\ref*{reg:#1}}} 
\newcommand{\AlgLine}[2]{\hyperref[alg:#1]{line~\ref*{line:#2} of Algorithm~\ref*{alg:#1}}}
\newcommand{\AlgLines}[3]{\hyperref[alg:#1]{lines~\ref*{line:#2}--\ref*{line:#3} of Algorithm~\ref*{alg:#1}}}
\else
\newcommand{\Sec}[1]{{Section~\ref{sec:#1}}} 
\newcommand{\App}[1]{{Appendix~\ref{sec:#1}}} 
\newcommand{\Eqn}[1]{{(\ref{eq:#1})}} 
\newcommand{\Part}[1]{{(\ref{part:#1})}} 
\newcommand{\Fig}[1]{{Figure~\ref{fig:#1}}} 
\newcommand{\Tab}[1]{{Table~\ref{tab:#1}}} 
\newcommand{\Thm}[1]{{Theorem~\ref{thm:#1}}} 
\newcommand{\Lem}[1]{{Lemma~\ref{lem:#1}}} 
\newcommand{\Prop}[1]{{Property~\ref{prop:#1}}} 
\newcommand{\Cor}[1]{{Corollary~\ref{cor:#1}}} 
\newcommand{\Def}[1]{{Definition~\ref{def:#1}}} 
\newcommand{\Alg}[1]{{Algorithm~\ref{alg:#1}}} 
\newcommand{\Ex}[1]{{Example~\ref{ex:#1}}} 
\newcommand{\As}[1]{{Assumption~\ref{as:#1}}} 
\newcommand{\Reg}[1]{{R~\ref{reg:#1}}} 
\newcommand{\AlgLine}[2]{{line~\ref{line:#2} of Algorithm~\ref{alg:#1}}}
\newcommand{\AlgLines}[3]{{lines~\ref{line:#2}--\ref{line:#3} of Algorithm~\ref{alg:#1}}}
\fi

\begin{document}

\def\spacingset#1{\renewcommand{\baselinestretch}%
	{#1}\small\normalsize} \spacingset{1}
	
\newcommand\independent{\protect\mathpalette{\protect\independenT}{\perp}}
\def\independenT#1#2{\mathrel{\rlap{$#1#2$}\mkern2mu{#1#2}}}

\if0\blind
{
	\title{\bf High-Dimensional Sparse Single-Index Regression Via Hilbert-Schmidt Independence Criterion }
	\author{Runxiong Wu\thanks{Runxiong Wu (E-mail:\href{mailto:11930643@mail.sustech.edu.cn}{11930643@mail.sustech.edu.cn}) is a Master student, Department of Statistics and Data Science, Southern University of
			Science and Technology, Shenzhen, China. }, \,
	Chang Deng\thanks{Chang Deng (E-mail: \href{mailto: changdeng@uchicago.edu}{changdeng@uchicago.edu}) is a Master student, Computational and Applied Mathematics, University of Chicago, Chicago, US. }\, and		
		Xin Chen\thanks{ Xin Chen (E-mail: \href{mailto:chenx8@sustech.edu.cn}{chenx8@sustech.edu.cn}) is an Associate Professor,
			Department of Statistics and Data Science, Southern University of
			Science and Technology, Shenzhen, China.}
                 \\}
	\date{}
	\maketitle
} \fi

\if1\blind
{
	\bigskip
	\bigskip
	\bigskip
	\begin{center}
		{\LARGE\bf Title}
	\end{center}
	\medskip
} \fi

\bigskip
\begin{abstract}
Hilbert-Schmidt Independence Criterion (HSIC) has recently been used in the field of single-index models to estimate the directions. Compared with some other well-established methods, it requires relatively weaker conditions. However, its performance has not yet been studied in the high-dimensional scenario, where the number of covariates is much larger than the sample size. In this article, we propose a new efficient sparse estimate in HSIC based single-index model. This new method estimates the subspace spanned by the linear combinations of the covariates directly and performs variable selection simultaneously. Due to the non-convexity of the objective function, we use a majorize-minimize approach together with the linearized alternating direction method of multipliers algorithm to solve the optimization problem. The algorithm does not involve the inverse of the covariance matrix and therefore can handle the large $p$ small $n$ scenario naturally. Through extensive simulation studies and a real data analysis, we show our proposal is efficient and effective in the high-dimensional setting. The $\verb|Matlab|$ codes for this method are available online.
\end{abstract}

\noindent%

\vfill

\newpage
\spacingset{1.5} 


\section{Introduction}
\label{sec:introduction}
Let $Y\in\bbR$ be an univariate response and $\X\in\bbR^p$ be a $p\times 1$ predictor, the single-index model as a practically useful generalization of the classical linear regression model considers the following problem
\begin{equation}\label{eqn1.1}
Y=g(\be^{\top}\X,\epsilon),
\end{equation}
where $\be$ is a $p\times 1$ vector, $\epsilon$ is an unknown random error independent of $\X$, and $g$ is a link function. Let $\mbox{span}(\be)$ denote the subspace spanned by $\be$, the goal of the single-index model is to estimate $\mbox{span}(\be)$ without specifying or estimating the link function $g$. To our best knowledge, \cite{li1989regression} firstly studied this problem and proposed to estimate the $\mbox{span}(\be)$ under the linear condition that $E(\X|\be^{\top}\X)$ is a linear function of $\be^{\top}\X$. This linear condition applies to the marginal distribution of $\X$ and is common in regression modeling.

Later, \cite{cook1994interpretation,cook2009regression} introduced sufficient dimension reduction (SDR) expanding the concept of the single-index model. It aims to find the minimal subspace $ \cS \subseteq \bbR^p$ such that $Y\independent \X|P_{\cS}\X$, where $\independent$ stands for independence and $P_{\cS}$ stands for the projection operator to the subspace $\cS$. Under mild conditions \citep{cook1996graphics,yin2008successive}, such a subspace exists and is unique. We call it the central subspace, denote it by $\cS_{Y|\X}$ and call its dimension $d=\mbox{dim}(\cS_{Y|\X})$, which is often far less than $p$. When the central subspace is one dimensional or in other words $d=1$, the caused regression problem is the single-index model (\ref{eqn1.1}). There are many methods proposed to estimate the central subspace \citep{li1991sliced,cook1991sliced,xia2002adaptive,cook2005sufficient,zhu2006fourier,li2007directional,wang2008sliced,cook2009likelihood,zeng2010integral,yin2011sufficient,ma2012semiparametric}. For a comprehensive list of references about SDR methods, please refer to \cite{ma2009likelihood}.

Unfortunately, one drawback of the dimension reduction methods mentioned above is that the estimated linear combinations still contain all the original predictors, which often makes it difficult to interpret the extracted components. To improve interpretability, numerous attempts have been made to perform variable selection and dimension reduction simultaneously, including \cite{cook2004testing,ni2005a,li2005model,li2007sparse,li2008sliced,chen2010coordinate}. These methods perform well when the number of covariates $p$ is less than the sample size denoted by $n$, but don't work under the scenario $p>n$. To tackle the difficulty, \cite{yin2015sequential} proposed sequential procedures in SDR and \cite{lin2018on}  proposed high-dimensional sparse sliced inverse regression (SIR). Moreover, \cite{wang2018estimating} introduced a reduced-rank regression method for estimating the sparse directions, and \cite{tan2018convex} proposed a convex formulation for fitting sparse SIR in high dimensions. Other recent high-dimensional SDR methods can be seen in \cite{qian2019sparse} and \cite{tan2020sparse}.

In this article, following the work of \cite{zhang2015direction} and \cite{tan2018convex}, we develop a new approach using Hilbert-Schmidt Independence Criterion (HSIC) for single-index models. The proposed method can handle the scenario $p>n$ and require the weakest conditions among the existing high-dimensional sparse SDR methods. The key idea is to formulate the HSIC based single-index model in a form of estimating the orthogonal projection $\be\be^{\top}$ onto the subspace span($\be$) rather than span($\be$), with the constraints of the nuclear norm and the operator norm to relax the normalization constraint. Moreover, our proposal uses a lasso penalty on the orthogonal projection $\be\be^{\top}$ to encourage the estimated solution to be sparse. To sum up, the main contributions of our work are as follows. First, our method extends the HSIC-based single-index regression \citep{zhang2015direction} to a sufficient variable selection method. Since it does not involve the inversion of the sample covariance matrix, it can naturally handle a large $p$ small $n$ situation. Second, motivated by the majorization-minimization principle, we design a fast and efficient algorithm to solve the problem. The objective function of our method is non-linear, so the algorithm in this article is more complicated and tricky than the algorithm in \cite{tan2018convex}. Third, \cite{tan2018convex} proposed a cross-validation scheme based on the idea of \cite{cook2008principal} to select the tuning parameters. Their method requires that the distribution of $\X|Y$ follows normal distribution, while our method apply a kernel method to estimate the link function which perfectly avoid this assumption. Last but not least, we can easily extend our method to situations where the response is multivariate.

The article is organized as follows. Section 2 reviews the background of HSIC-based single-index method and Section 3 details our proposed method. In Section 4, we conduct extensive simulation studies and a real data analysis. A short conclusion and some technical proofs are provided in Section 5 and Appendix.

The following notations will be used in our exposition. Let $\|\cdot\|$ denote the Euclidean norm of a vector in the corresponding dimension, $\|\cdot\|_1$ denote the sum of elementwise absolute values, $\|\cdot\|_{\rm F}$ denote the Frobenius norm of a matrix, and $\|\cdot\|_{*}$ denote the nuclear norm of a matrix. $P_{{\bm\eta}(\bSigma)}={\bm\eta}({\bm\eta}^{\top} \bSigma{\bm\eta})^{-1}{\bm\eta}^{\top}\bSigma$ denotes the projection operator which projects onto span(${\bm\eta}$) relative to the inner product $\langle {\bf a},{\bf b}\rangle = {\bf a}^{\top} \bSigma {\bf b}$ and                                                                                                                                                                                                                                                                                                                                          $Q_{{\bm\eta}(\bSigma)}=\I-P_{{\bm\eta}(\bSigma)}$, where $\I$ is the identity matrix. The trace of a matrix ${\bf A}$ is tr({\bf A}) and the Euclidean inner product of two matrices ${\bf A}, {\bf B}$, is $\langle{\bf A},{\bf B} \rangle$=tr$({\bf A}^{\top}{\bf B})$. $\bI_{(a>0)}$ is the indicator function and $\lambda_{\max}( \cdot)$ is the largest eigenvalue of a matrix.

\section{Overview of HSIC-based Single-Index Regression}
\label{sec:meth}
\cite{gretton2005measuring,gretton2007kernel,gretton2009discussion} proposed an independence criterion termed the Hilbert-Schmidt Independence Criterion to detect statistically significant dependence between two random variables. For univariate $X$ and $Y$, HSIC denoted by $H(X,Y)$ has a population expression
\begin{equation}\label{eqn2.1}
\begin{split}
H(X,Y)=&E\left[ K(X-X')L(Y-Y')\right]+E\left[ K(X-X')\right]E\left[L(Y-Y')\right]\\
&-2E\left\{  E\left[ K(X-X')|X \right]  E\left[ L(Y-Y')|Y\right]  \right\},
\end{split}
\end{equation}
where $X'$ and $Y'$ denote independent copies of $X$ and $Y$, and $K(\cdot)$ and $L(\cdot)$ are
positive definite kernel functions. The definition of HSIC exists when the various expectations over the kernels are finite, which is true as long as the kernels $K(\cdot)$ and $L(\cdot)$ are bounded. One often used kernel is a Gaussian kernel \citep[see][]{kankainen1995consistent}, i.e.,
\begin{equation*}
K:=\exp\left( \frac{-\|X-X'\|^2}{2\sigma_X^2} \right) \mbox{ and } L:= \exp\left( \frac{-\|Y-Y'\|^2}{2\sigma_Y^2} \right).
\end{equation*}
Moreover, \cite{feuerverger1993consistent} showed that the statistic is equivalent to the characteristic function-based statistic when the Gaussian kernel choice is adopted.
Throughout the article, we present our method using the Gaussian kernel, however, our method can be extended to other kernel choices without much issue.

According to \cite{gretton2005kernel}, HSIC equals $0$ if and only if two random variables are independent, which makes it possible for its application in the field of SDR. Indeed, under a mild condition, \cite{zhang2015direction} showed that solving $(\ref{eqn2.2})$ with respect to a $p\times 1$ vector $\be$ would yield a basis of $\cS_{Y|\X}$, or in other words, the single-index direction:
\begin{equation}\label{eqn2.2}
\underset{\be^{\top}{\bf\Sigma}\be=1}{\max}\, H(\be^{\top}\X,Y),
\end{equation}
where ${\bSigma}$ denotes the covariance matrix of $\X$. Note that solving $(\ref{eqn2.2})$ may not have a unique solution in terms of $\be$, but we are interested only in span($\be$), which is unique as shown in the following proposition.
\begin{proposition}
\label{prop:2.1.}
Assume that the support of $\X \in\bbR^p$ is a compact set, and that ${\bm\eta}$ spans the central subspace such that ${\bm\eta}^{\top}\bSigma{\bm\eta}=1$. If $P^{\top}_{{\bm\eta}(\bSigma)}\X \independent Q^{\top}_{{\bm\eta}(\bSigma)}\X $, then any result $\be$ of solving $(\ref{eqn2.2})$ satisfies span($\be$)=span(${\bm\eta}$).
\end{proposition}

Let $\left( \X,\Y \right)=\left\{ (\X_i,Y_i): i=1,\ldots,n  \right\}$ be a random sample of $n$ i.i.d. random vectors $(\X,Y)$, and $\hat{\bSigma}$ and $\hat{\sigma}_Y$ be the sample covariance matrix and sample variance of $\X$ and $Y$, respectively. The corresponding sample version of $H(\be^{\top}\X,Y)$, denoted by $H_n(\be^{\top}\X,\Y)$, is a sum of three U-statistics \citep[see][]{serfling2009approximation,gretton2007kernel}:
\begin{equation}\label{eqn2.3}
H_n(\be^{\top}\X,\Y)=\frac{1}{n^2}\sum_{i,j=1}^{n} K_{ij}(\be)L_{ij}-\frac{2}{n^3} \sum_{i,j,k=1}^{n}K_{ij}(\be)L_{ik}+\frac{1}{n^4}\sum_{i,j,k,l=1}^{n} K_{ij}(\be)L_{kl},
\end{equation}
where
\begin{equation*}
K_{ij}(\be):=\exp\left(  \frac{-(\be^{\top}(\X_i-\X_j) )^2}{2\be^{\top}\hat{\bSigma}\be} \right) \mbox{ and } L_{kl}:=\exp\left(\frac{-\|Y_k-Y_l\|^2}{2\hat{\sigma}_{Y}^2}  \right).
\end{equation*}
In later sections, we will utilize the equivalent form \citep[see][]{gretton2007kernel,wu2019mm}, obtained by replacing the U-statistics with V-statistics
\begin{equation}\label{eqn2.4}
H_n(\be^{\top}\X,\Y)=\frac{1}{n^2}\mbox{tr}(\K\H\L\H)=\frac{1}{n^2}\sum_{i,j=1}^{n} K_{ij}(\be)  \tilde{L}_{ij}
\end{equation}
rather than Equation $(\ref{eqn2.3})$, where $\K$ and $\L$ are the $n\times n$ matrix with entries $K_{ij}(\be)$ and $L_{ij}$ respectively, $ \displaystyle \H=\I-\frac{1}{n}\1\1^{\top} $, and $\1$  is a $n\times 1$ vector of ones. Here, $\tilde{L}_{ij}$ denotes the $(i,j)$-th entry of the product matrix $\H\L\H$. The estimator of a basis for the central subspace $\cS_{Y|\X}$ is
\begin{equation}\label{eqn2.5}
{\bm\eta}_{n}=\underset{ \be^{\top}\hat{\bSigma}\be=1}{\arg\max}\, H_{n}(\be^{\top}\X,\Y).
\end{equation}
Then, the central subspace is estimated as span(${\bm\eta}_{n}$) and the sufficient dimension reduced variable is ${\bm\eta}_{n}^{\top}\X$. The following proposition characterizes the asymptotic properties of the estimator ${\bm\eta}_{n}$.
\begin{proposition}
\label{prop:2.2.}
Under the assumptions in Proposition 1, if ${\bm\eta}_{n}=\underset{ \be^{\top}\hat{\bSigma}\be=1}{\arg\max}\, H_{n}(\be^{\top}\X,\Y)$, then ${\bm\eta}_{n}$ converges in probability to $c{\bm\eta}$ as $n \to \infty$, where $c=1$ or $c=-1$. Furthermore, under some other regularity conditions, $\sqrt{n}({\bm\eta}_{n}-c{\bm\eta})\to N(0,{\bf V}_{11})$, where ${\bf V}_{11}$ is a covariance matrix.
\end{proposition}
For details about the regularity conditions and the specific form of ${\bf V}_{11}$, please refer to \citet{zhang2015direction} and its online supplementary material.

\section{Method}
\label{sec: sparse estimate}
\subsection{Problem Formulation}
Let $\bPi=\be\be^{\top}$, the HSIC-based single-index regression procedure (\ref{eqn2.5}) can be rewritten as the following minimization problem:
\begin{equation}\label{eqn3.1}
\begin{split}
&\underset{ \bPi\in \cM}{\rm min}\, -\frac{1}{n^2}\sum_{i,j=1}^{n} \exp\left( -\frac{ \langle  \bPi,\Z_{ij} \rangle }{2} \right) \tilde{L}_{ij}, \\
&\mbox{   s.t. }\quad \hat{\bSigma}^{1/2} \bPi\hat{\bSigma}^{1/2} \in \cB,
\end{split}
\end{equation}
where $\Z_{ij}=(\X_i-\X_j)(\X_i-\X_j)^{\top}$, $\cB=\left\{ \hat{\bSigma}^{1/2} \bPi\hat{\bSigma}^{1/2}: \be^{\top}\hat{\bSigma}\be=1   \right\}$, and $\cM$ is the set of $p \times p$ symmetric semi-definite positive matrices. In this new formulation, our focus is changed to directly estimate the orthogonal projection $\bPi$ onto the subspace instead of estimating the basis $\be$.

In high dimensional SDR, it is often true that only a few elements of $\X$ are informative and we would like to select these variables only. To achieve this goal, \cite{tan2018convex} introduces the notion of subspace sparsity and imposes a lasso penalty on all elements of $\bPi$ to encourage such sparsity. Moreover, they utilize the nuclear norm and the spectral norm to relax the constraint. Following the work of them, we propose the sparse estimate by solving
\begin{equation}\label{eqn3.2}
\begin{split}
&\underset{ \bPi\in \cM}{\rm min}\, -\frac{1}{n^2}\sum_{i,j=1}^{n} \exp\left( -\frac{ \langle  \bPi,\Z_{ij} \rangle }{2} \right) \tilde{L}_{ij} +\lambda \|\bPi\|_{1}, \\
&\mbox{   s.t. }\quad {\rm tr}(\hat{\bSigma}^{1/2} \bPi\hat{\bSigma}^{1/2} )\leq1,
\end{split}
\end{equation}
where $\lambda$ is a tunning parameter. Note that we only consider the dimension of the central subspace to be 1, so there is no need to impose spectral norm constraint. More similar work can be seen in sparse principal component analysis, canonical correlation analysis, and sliced inverse regression \citep{vu2013fantope,gao2017sparse,tan2018sparse,tan2018convex,tan2020sparse}. In addition, when the kernel is the product kernel, we can naturally extend the method to settings where the response is multivariate. That is, for a $q$-dimensional response $\Y=(Y_1,\ldots,Y_q)^{\top}$, we use the product kernel:
\begin{equation}\notag
L:=\prod_{i=1}^{q}\exp\left( \frac{-|Y_i-Y_i^{'}|^2}{2\sigma_{Y_i}^2} \right),
\end{equation}
where $\Y'=(Y_1^{'},\ldots,Y_q^{'})^{\top}$ is an independent copy of $\Y$.

\subsection{Computation}
In this subsection, we propose an efficient optimization algorithm for solving the problem $(\ref{eqn3.2})$. Let $f(\bPi)$ denote the objective function of the problems $(\ref{eqn3.1})$. Although $f(\bPi)$ is not convex, it is differentiable and has Lipschitz continuous gradient over the bounded convex set. We state properties about the objective function $f(\bPi)$ in the following proposition.
\begin{proposition}
\label{prop:3.1.}
$f(\bPi)$ is differentiable and its derivative function is
\begin{equation}
\label{eqn3.3}
\nabla f(\bPi)=\frac{1}{2n^2} \sum_{i,j=1}^{n} \exp\left(-\frac{ \langle \bPi, \Z_{ij} \rangle    }{2} \right)\tilde{L}_{ij}\Z_{ij},
\end{equation}
or equivalently,
\begin{equation}
\label{eqn3.4}
\nabla f(\bPi)=\frac{1}{n^2} \X^{\top}\left(  {\rm diag}({\bf C}{\bm 1}_{n})-  {\bf C}\right)\X,
\end{equation}
where ${\bf C}$ is a $n\times n$ matrix with the entry $\displaystyle c_{ij}=\exp\left(-\frac{ \langle \bPi, \Z_{ij} \rangle    }{2} \right)\tilde{L}_{ij}$ and $\X=[\X_1,\ldots,\X_n]^{\top}$. Moreover, $\nabla f(\bPi)$ is Lipschitz over the set $\cD=\left\{ \bPi\in \cM,  {\rm tr}(\hat{\bSigma}^{1/2} \bPi\hat{\bSigma}^{1/2} )\leq 1 \right\}$.
\end{proposition}
\noindent We prove the Proposition $\ref{prop:3.1.}$ in the Appendix.
\begin{remark}
It is worth noting that we would like to use the expression form $(\ref{eqn3.4})$ instead of $(\ref{eqn3.3})$ when actually calculating the derivative function $\nabla f(\bPi)$. Plus, the Lipschitz continuity property of $f(\bPi)$
motivates us to design a method for performing the optimization in this work from the viewpoint of the majorization-minimization principle \citep{lange2000optimization,hunter2004tutorial}.
\end{remark}

Since the objective function $ f(\bPi)$ has a Lipschitz continuous gradient over the bounded set $\cD$, there exists a positive constant $L < \infty$ such that
\begin{equation}
\label{eqn3.5}
f(\bPi)\leq f(\tilde{\bPi})+\langle \bPi-\tilde{\bPi}, \nabla f(\tilde{\bPi}) \rangle+\frac{L}{2} \|\bPi-\tilde{\bPi}\|_{\rm F}^2,
\end{equation}
for all $\bPi\in\cD$ and $\tilde{\bPi}\in\cD$. Thus,  the right hand side of ($\ref{eqn3.5}$) is a majorizing
function of $f(\bPi)$ at $\bPi$ (i.e., the right hand side of ($\ref{eqn3.5}$) is greater than or equal to $f(\bPi)$  for all $\bPi\in\cD$ with equality when $\bPi=\tilde{\bPi}$). This suggests the following majorize-minimize (MM) iteration to solve the problem ($\ref{eqn3.2}$):
\begin{eqnarray}
\bPi^{(r+1)}&=&\underset{\bPi\in\cD}{\arg\min}\, \left\{ f(\bPi^{(r)})+\langle \bPi-\bPi^{(r)}, \nabla f(\bPi^{(r)}) \rangle+\frac{L}{2} \|\bPi-\bPi^{(r)}\|_{\rm F}^2+\lambda\|\bPi\|_1 \right\},\notag\\
 &=& \underset{\bPi\in\cD}{\arg\min} \, \frac L2\Big\|\bPi-\left[ {\bf\Pi}^{(r)}-\frac 1L \nabla f({\bf\Pi}^{(r)}) \right]  \Big\|_{\rm F}^2 + \lambda \|\bPi\|_1 ,\label{eqn3.6}
\end{eqnarray}
where $\bPi^{(r+1)}$ and $\bPi^{(r)}$ are the  $(r+1)$-th and $r$-th iterates of the optimization variable corresponding to $\bPi$, respectively. By the property $(\ref{eqn3.5})$, we can easily obtain
\begin{equation}\notag
f(\bPi^{(r+1)})+ \lambda \|\bPi^{(r+1)}\|_{1} \leq  f(\bPi^{(r)})+\lambda \|\bPi^{(r)}\|_{1} \mbox{ for all } r,
\end{equation}
which means that iterates generated from the algorithm are guaranteed to monotonically decrease the objective function value.
\cite{hunter2004tutorial} showed the sequence $\left\{ \bPi^{(r)} \right\}_{r\geq 0}$ obtained by the iterative formula $(\ref{eqn3.6})$ converges to a critical point of the problem $(\ref{eqn3.2})$. The MM algorithm is a well-applicable and simple algorithmic framework for solving such problems. The key challenge in making the proposed algorithm efficient numerically lies in solving the subproblem $(\ref{eqn3.6})$.

The subproblem $(\ref {eqn3.6})$ is a quadratic problem with the convex constraint, so any local minimum can be guaranteed to be a global minimum. We employ the linearized alternating direction method of multipliers algorithm \citep[L-ADMM,][]{zhang2011Bregman, wang2012LADMM, yang2013LADMM} to solve it. This algorithm can allow us to tackle the difficult caused by the interaction between the penalty term and the constraints. We give the derivation details of solving the subproblem $(\ref {eqn3.6})$ through this algorithm in the Appendix. In practice, we find that this algorithm can solve the subproblem quite efficiently.

Algorithm $\ref{alg1}$ presents the entire algorithm flow we use to solve the problem $(\ref{eqn3.2})$. It has two loops: an outer loop in which the MM algorithm approximates the original problem $(\ref{eqn3.2})$ iteratively by a series of
convex relaxations, and an inner loop in which the linearized alternating direction method of multipliers algorithm is used to solve each convex relaxation $(\ref{eqn3.6})$. In the inner loop, the update of $\bPi$ is performing soft-thresholding and the update of $\H$ is a projection operator which needs to compute a singular value decomposition, and modify the obtained singular values with a monotone piecewise linear function. For specific details about the projection operator, please refer to the Proposition \ref{proA.1} in the Appendix. \verb|Matlab| codes implementing the method are available at \url{https://github.com/runxiong-wu/sHSIC}.

\begin{algorithm}[!h]
	\caption{MM Algorithm for Solving ($\ref{eqn3.2}$)}
	\label{alg1}
	\KwIn{$\X, Y$, the tuning parameter $\lambda$, the Lipschitz constant $L$, the L-ADMM
		parameters $\rho > 0$ and $\tau=4\rho \lambda^2_{\rm max}(\hat{\bSigma}).$}
	Initialize ${\bf\Pi}^{(0)}\in\cM$ and $\H^{(0)}=\hat{\bf\Sigma}^{1/2}{\bf\Pi}^{(0)}\hat{\bf\Sigma}^{1/2}$\;
	\Repeat( $r=0,1,2,\ldots$ ){stopping criterion met}{
		Initialize primal variables ${\bf\Pi}_{0}= {\bf\Pi}^{(r)}, \H_0=\H^{(r)},$ and dual variable ${\bf\Gamma}_0=\0$\;
		\Repeat( $j=0,1,2,\ldots$){stopping criterion met}{
			  temp $\gets \displaystyle \frac{L}{L+\tau} \left[ {\bf\Pi}^{(r)}-\frac{\nabla f({\bf\Pi}^{(r)})}{L}\right]$\;
			  temp $\gets$ temp $ + \displaystyle \frac{\tau}{L+\tau} \left[ {\bf\Pi}_{j}-\frac{\rho}{\tau}\hat{\bf\Sigma}{\bf\Pi}_{j}\hat{\bf\Sigma}+\frac{\rho}{\tau}\hat{\bf\Sigma}^{1/2}(\H_j-{\bf\Gamma}_{j})\hat{\bf\Sigma}^{1/2} \right] $\;
			  $\displaystyle {\bf\Pi}_{j+1} \gets {\rm Soft}\left( \mbox{temp},\frac{\lambda}{L+\tau}  \right)$, i.e., soft-thresholding elementwise\;
			  $\H_{j+1} \gets P_{\cF}(\hat{\bf\Sigma}^{1/2}{\bf\Pi}_{j+1}\hat{\bf\Sigma}^{1/2}+{\bf\Gamma}_{j}) $, see Proposition \ref{proA.1} in the Appendix\;
			  ${\bf\Gamma}_{j+1} \gets {\bf\Gamma}_{j}+\hat{\bf\Sigma}^{1/2} {\bf\Pi}_{j+1}\hat{\bf\Sigma}^{1/2}-\H_{j+1} $\;
		}
	    ${\bf\Pi}^{(r+1)}\gets {\bf\Pi}_{j+1},\H^{(r+1)}\gets\H_{j+1}, {\bf\Gamma}^{(r+1)}\gets {\bf\Gamma}_{j+1}$\;		
	}
   	\KwOut{$\hat{\be}=\mbox{the most top eigenvector of } {\bf\Pi}^{(r+1)}.$}
\end{algorithm}

\subsection{Tuning Parameter Selection}
The tuning parameter $\lambda$ in our proposed method determines the sparsity level of the estimate. \cite{tan2018convex} proposed a cross-validation approach based on the framework of principal fitted components \citep[PFC,][]{cook2008principal} to select the corresponding sparsity tuning parameter. However, the PFC method requires that the distribution of $ \X | Y $ should be normally distributed, which may not be suitable in the real application. To avoid the assumption, we use the Nadaraya-Watson kernel method to estimate the conditional expectation $E(Y|\X)$. Let $\hat{\bPi}$ be an estimate of the orthogonal projection $\bPi$, the sufficient dimension direction estimator $\hat{\be}$ is estimated by the top eigenvector of $\hat{\bPi}$. Given a new data $\X^*$, the Nadaraya-Watson kernel estimator of conditional mean $E(Y|\X=\X^*)$ is
\begin{equation}\label{eqn3.7}
\hat{E}(Y|\X=\X^*)=\sum_{i=1}^{n} \frac{ K_h(\hat{\be}^{\top}(\X^*-\X_i)) }{ \sum_{j=1}^{n} K_h(\hat{\be}^{\top}(\X^*-\X_j)) } Y_i,
\end{equation}
where $K_h$ is a kernel with a bandwidth $h$. 
In this article, we use a Gaussian kernel and take the leave-one-out estimate for bandwidth selection. Note that there is a trick to compute the cross-validation function with a single fit. This trick vastly reduces the computational complexity, at the price of the increasing memory consumption. For specific details, please refer to \cite{fan1996local}.

We then construct an M-fold cross-validation procedure based on $(\ref{eqn3.7})$ to select the tuning parameter $\lambda$. Suppose $C_1,\ldots,C_M$ are $M$ equally sized and mutually disjoint subsamples of the whole dataset. The cross-validation procedure utilizes each single subsample be the test data, and the remaining $M-1$ subsamples be the training data. For each fixed tuning parameter $\lambda$, the corresponding overall prediction error is computed as $\sum_{m=1}^{M} \sum_{i\in C_m} \left\{   Y_i- \hat{E}(Y|\X=\X_i)\right\}^2/(M|C_m|)$ where $|C_m|$ denotes the cardinality of the set $C_m$. Finally, we choose the tuning parameter which minimizes the prediction error.

\section{Numerical Study}
\subsection{Simulations}
\label{sec:numerical}
In this section, we compare the performance of our proposed method with the most competitive high-dimensional sparse SDR approach \citep{tan2018convex} under various simulation settings. We use two measures: the true positive rate (TPR) and the false positive rate (FPR), to assess how well the methods select variables. In particular, TPR is defined as the proportion of active predictors that are correctly identified while FPR is defined as the proportion of irrelevant predictors that are falsely identified. An estimate with a bigger TPR and a smaller FPR is better. Furthermore, we calculate the absolute correlation coefficient (corr) between the true sufficient predictor and its estimate to evaluate accuracy of the methods. The larger the absolute correlation coefficient, the better the estimate. For each study, we repeat 200 times.

\begin{table}[!h]	
	\centering
	\caption{Summary of the simulation studies. The mean, averaged over 200 datasets, are reported. All entries are multiplied by 100.}
	\label{tabel3}
	\resizebox{\textwidth}{!}{
		\begin{threeparttable}		
			\begin{tabular}{cccccccccc}
				\toprule
				\multirow{2}{*}{}&\multirow{2}{*}{}&\multicolumn{4}{c}{ $n=100,p=150$ }&\multicolumn{4}{c}{ $n=200,p=150$ }\cr			
				\cmidrule(lr){3-6} \cmidrule(lr){7-10}
				& & Study 1 & Study 2 & Study 3& Study 4 & Study 1 & Study 2 & Study 3 & Study 4\cr
				\midrule
Our proposed method&TPR &73.8 &99.3&91.1&78.8& 88.8 &100&98.0&94.7 \cr
				   &FPR &3.3 &0.8 &4.9 &0.9 & 1.3 &0.4 &1.2 &0.6 \cr	
				   &corr&70.8 &95.3&84.3&82.5& 83.7 &98.3&95.9&87.9 \cr
Tan et al. (2018)&TPR &76.7 &98.7&66.3&43.8&97.8  &100 &67.9  &59 \cr
				&FPR &3.6  &1.4 &37.5&8.9 &2.6  &1.1 &2.6&0.7 \cr	
				&corr&69.6 &91.9&32.1&48.8&89.9&97.5&64&71.8 \cr
				\bottomrule
			\end{tabular}	
	\end{threeparttable}}
\end{table}

\begin{itemize}[topsep=0pt]
	\item[{\it Study 1.}] This model is a classic linear regression model from \cite{tan2018convex}:
	\begin{equation}\notag
	Y=\beta^{\top}\X+2\epsilon,
	\end{equation}
	where $\epsilon \sim N(0,1)$, $\X=(X_1,\ldots,X_p)^{\top}\sim N_{p}(\0,{\bf \Sigma})$ with ${\bf \Sigma}_{ij}=0.5^{|i-j|}$ for $1\leq i,j\leq p$, and $\X$ and $\epsilon$ are independent. In this study, the central subspace is spanned by the vector $\beta=(1,1,1,0,\ldots,0)^{\top}/\sqrt{3}$ with $p-3$ zero coefficients.
	
	\item[{\it Study 2.}] This model is a nonlinear regression model from \cite{yin2015sequential}:
	\begin{equation}\notag
	Y=1+\exp(\beta^{\top}\X)+\epsilon,
	\end{equation}
	where $\epsilon \sim N(0,1)$, $\X=(X_1,\ldots,X_p)^{\top}\sim N_{p}(\0,{\bf \Sigma})$ with ${\bf \Sigma}_{ij}=0.5^{|i-j|}$ for $1\leq i,j\leq p$, and $\X$ and $\epsilon$ are independent. In this study, the central subspace is spanned by the vector $\beta=(1,1,1,0,\ldots,0)^{\top}/\sqrt{3}$ with $p-3$ zero coefficients.
	
	\item[{\it Study 3.}] This model is from \cite{chen2018efficient}:
	\begin{equation}\notag
	Y=(\beta^{\top}\X+0.5)^2+0.5\epsilon,
	\end{equation}
	where $\epsilon \sim N(0,1)$, $\X=(X_1,\ldots,X_p)^{\top}\sim N_{p}(\0,{\bf \Sigma})$ with ${\bf \Sigma}_{ij}=0.5^{|i-j|}$ for $1\leq i,j\leq p$, and $\X$ and $\epsilon$ are independent. In this study, the central subspace is spanned by the vector $\beta=(1,1,1,1,0,\ldots,0)^{\top}/2$ with $p-4$ zero coefficients.
	
	\item[{\it Study 4.}] This model is a mean function model similar to \cite{zhang2015direction}:
	\begin{equation}\notag
	Y=\sin(\beta^{\top}\X)+0.2\epsilon,
	\end{equation}
	where $\epsilon \sim N(0,1)$. The predictor $\X=(X_1,\ldots,X_p)^{\top}$ is independent of $\epsilon$ and defined as follows: the last $p-1$ components $(X_2,\ldots,X_p)^{\top} \sim N_{p-1}(\0,{\bf \Sigma}) $ with ${\bf \Sigma}_{ij}=0.5^{|i-j|}$ for $1\leq i,j\leq p-1$ and the first component $X_1=|X_2+X_3|+0.1 \xi$, where $\xi$ is an independent standard normal random variable. In this study, the central subspace is spanned by the vector $\beta=(1,1,1,0,\ldots,0)^{\top}/\sqrt{3}$ with $p-3$ zero coefficients.
	
	\item[{\it Study 5.}] This model is a multivariate response model combining Study 1 and Study 3:
	\begin{equation}\notag
	\left\{
	\begin{aligned}
	Y_1&=\beta^{\top}\X+2\epsilon, \\
	Y_2&=(\beta^{\top}\X+0.5)^2+0.5\epsilon,
	\end{aligned}
	\right.
	\end{equation}
	where $\epsilon \sim N(0,1)$. The predictor $\X=(X_1,\ldots,X_p)^{\top}$ is independent of $\epsilon$ and defined as the same as the Study 3. In this study, $\beta=(1,1,1,1,0,\ldots,0)^{\top}/2$ with $p-4$ zero coefficients.
	
	\item[{\it Study 6.}] This model is a multivariate response model combining Study 3 and Study 4:
	\begin{equation}\notag
	\left\{
	\begin{aligned}
	Y_1&=(\beta^{\top}\X+0.5)^2+0.5\epsilon, \\
	Y_2&=\sin(\beta^{\top}\X)+0.2\epsilon,
	\end{aligned}
	\right.
	\end{equation}
	where $\epsilon \sim N(0,1)$. The predictor $\X=(X_1,\ldots,X_p)^{\top}$ is independent of $\epsilon$ and defined as the same as the Study 4. In this study, $\beta=(1,1,1,1,0,\ldots,0)^{\top}/2$ with $p-4$ zero coefficients.
	
\end{itemize}

Let $\hat{\bPi}$ be an estimator of the orthogonal projection $\bPi$, the sufficient dimension direction estimator $\hat{\be}$ is obtained by computing the top eigenvector of $\hat{\bPi}$. When computing the TPR and the FPR in practice, we truncated $\hat{\be}$ by zeroing out its entries whose magnitude is smaller than $10^{-4}$. For the method in \cite{tan2018convex}, we use Tan's code with the default parameter setting.

The simulation results from Study 1 to Study 4 are summarized in Table \ref{tabel3}. We can see that although our proposed method in Study 1 is slightly better than the method of \cite{tan2018convex} in terms of FPR, it is worse than \cite{tan2018convex} in general. This phenomenon is well explained by that the SIR method has the best performance in a classic linear model. In Study 2, our method outperforms the other method slightly in general. The performance of the method in \cite{tan2018convex} relies on the choices of the method-specific kernel matrix while our method does not have this limit. In Study 3, the conditional distribution is approximately symmetrical, which causes serious problem to the method of \cite{tan2018convex}. However, our method is still valid in this case. In Study 4, the linearity condition about $\X$ is destroyed while most of SDR methods require this condition. Thus, in such a case it is not surprising that our proposed method performs better than the rest method. In short, our proposed method performs very well across all the four studies in the high-dimensional setting. Studies 5 and 6 investigate the effect of our proposed method about variable selection in a multivariate response model. As far as we know, it seems no apparent competitor in such scenarios. The results are summarized in Table \ref{tabel4} and we can see our proposed method works fine even if the response is multivariate.

\begin{table}[!htbp]	
	\centering
	\caption{Summary of the simulation studies 5 and 6. The mean, averaged over 200 datasets, are reported. All entries are multiplied by 100.}
	\label{tabel4}
	\resizebox{\textwidth}{!}{
		\begin{threeparttable}		
			\begin{tabular}{cccccc}
				\toprule
				\multirow{2}{*}{}&\multirow{2}{*}{}&\multicolumn{2}{c}{ $n=100,p=150$ }&\multicolumn{2}{c}{ $n=200,p=150$ }\cr			
				\cmidrule(lr){3-4} \cmidrule(lr){5-6}
				& & Study 5 & Study 6  & Study 5 & Study 6\cr
				\midrule
				Our proposed method&TPR &99.8 &98.9&100.0&100.0 \cr
				&FPR &0.7 &2.7 &0.4 &1.7 \cr	
				&corr& 95.1 &92.5&98.2&95.2 \cr
				\bottomrule
			\end{tabular}	
	\end{threeparttable}}
\end{table}	

\subsection{Real Data Analysis}
In this part, we evaluate the performance of our proposed method in a real dataset about riboflavin (vitamin $\mbox{B}_2$) production with Bacillus subtilis, which is publicly available in the \verb|R| package \verb|hdi|. This dataset was analyzed by \cite{dezeure2015high}, \cite{hilafu2017sufficient}, and \cite{shi2020statistical}. It consists of a single real-valued response variable which is the logarithm of the riboflavin production rate and $p=4088$ predictors measuring the logarithm of the expression level of $4088$ genes. The purpose is to systematically search genomic features that contain sufficient
information for riboflavin production rate response prediction. We center the response and standardize all the covariates before
analysis.

The sample size $n=71$ is small compared with the covariate dimension $p=4088$. To handle the ultrahigh dimensionality, we preselect the most significant $100$ genes via the DC-SIS \citep{li2012feature}. Following the work of \cite{hilafu2017sufficient}, we split the data into a training set of 50 samples and a test set of 21 samples. The
training set is used to select features and estimate the central subspace. To evaluate the performance in the test data, we fit a linear model with the selected variables as predictors, rather than building a complex model. 
\begin{figure}[!htbp]
\centering
\subfigure[]{
	\begin{minipage}[t]{0.5\linewidth}
		\centering
		\includegraphics[width=\textwidth]{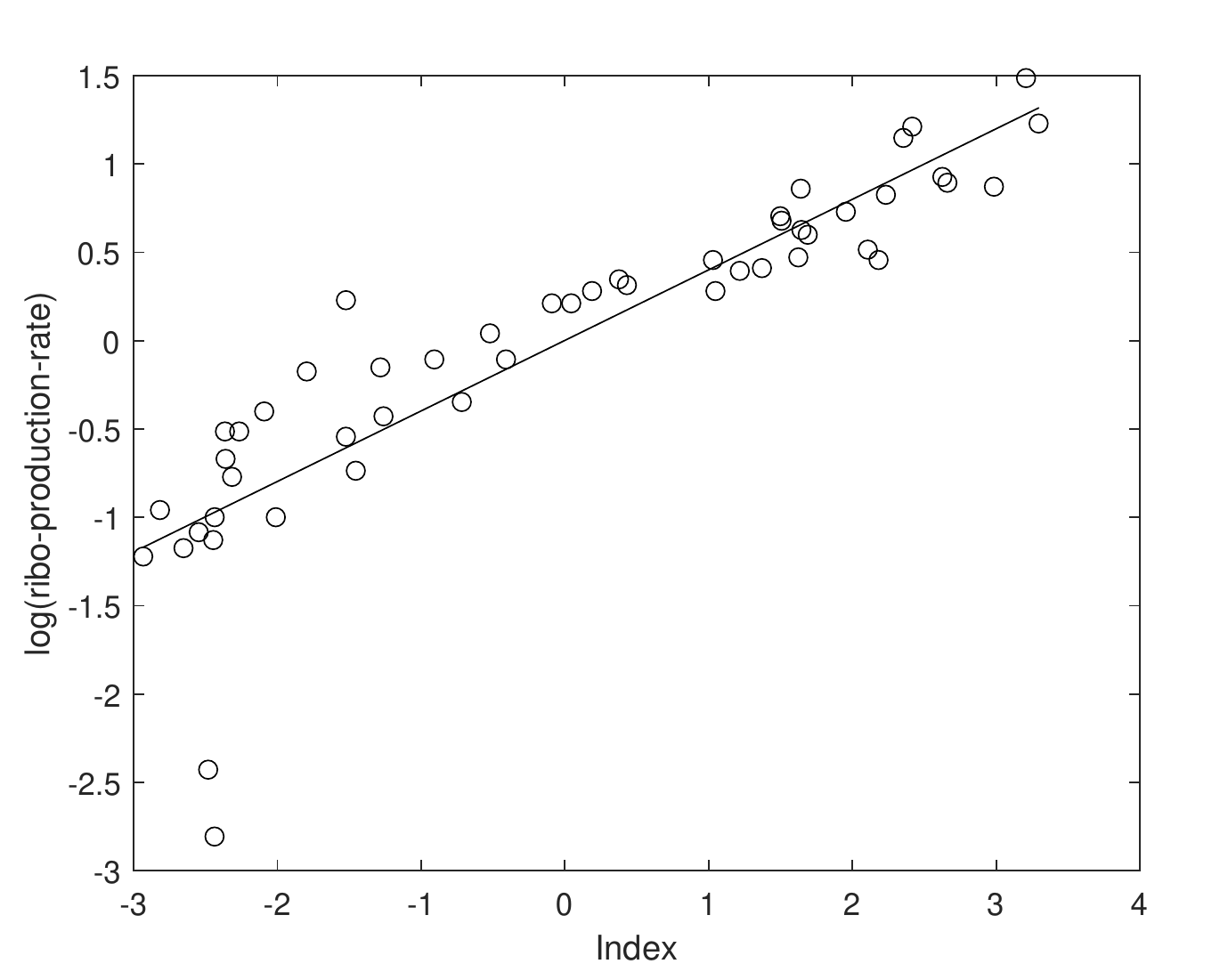}
	\end{minipage}%
}%
\subfigure[]{
	\begin{minipage}[t]{0.5\linewidth}
		\centering
		\includegraphics[width=\textwidth]{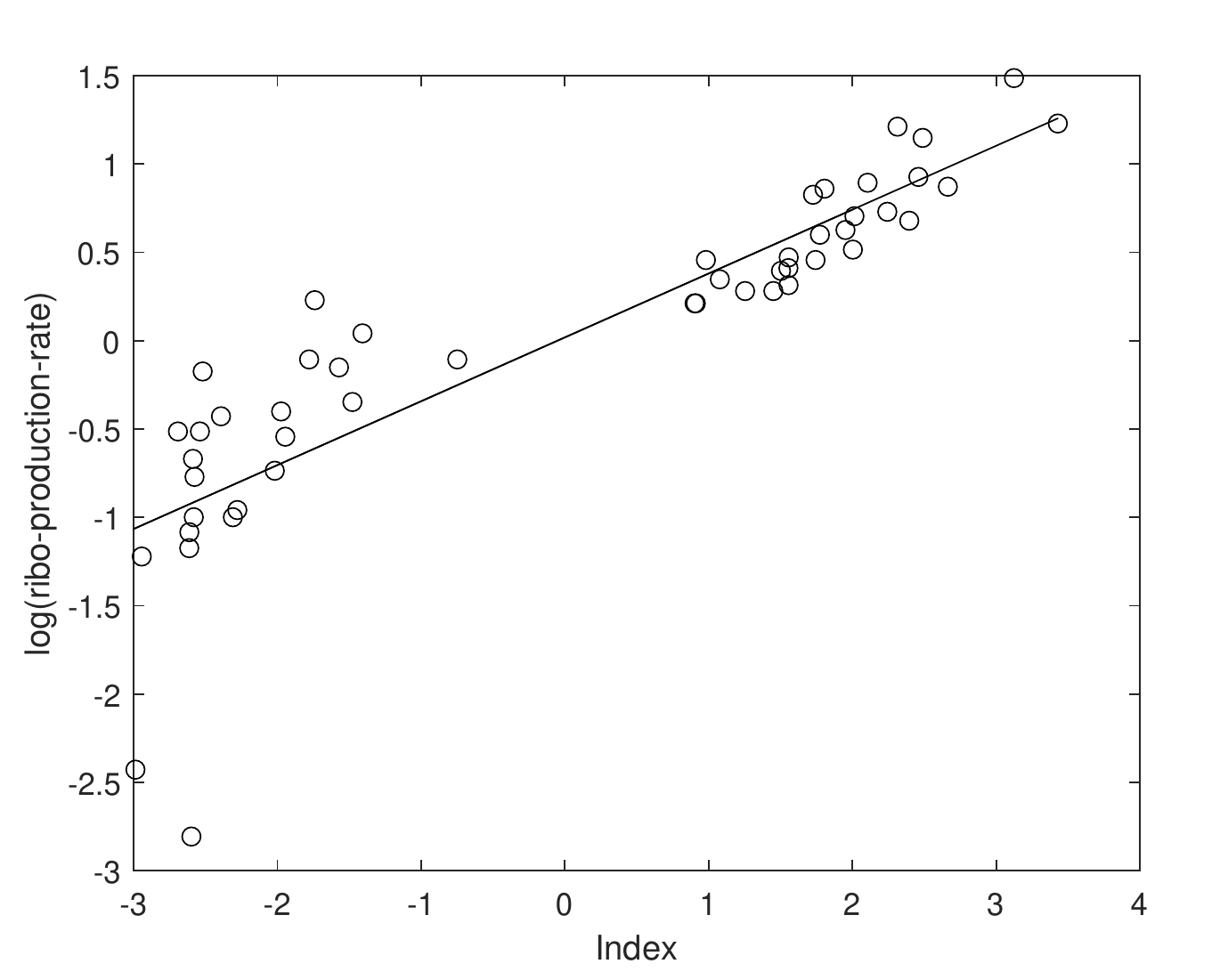}
	\end{minipage}%
}%
\\
\subfigure[]{
	\begin{minipage}[t]{0.5\linewidth}
		\centering
		\includegraphics[width=\textwidth]{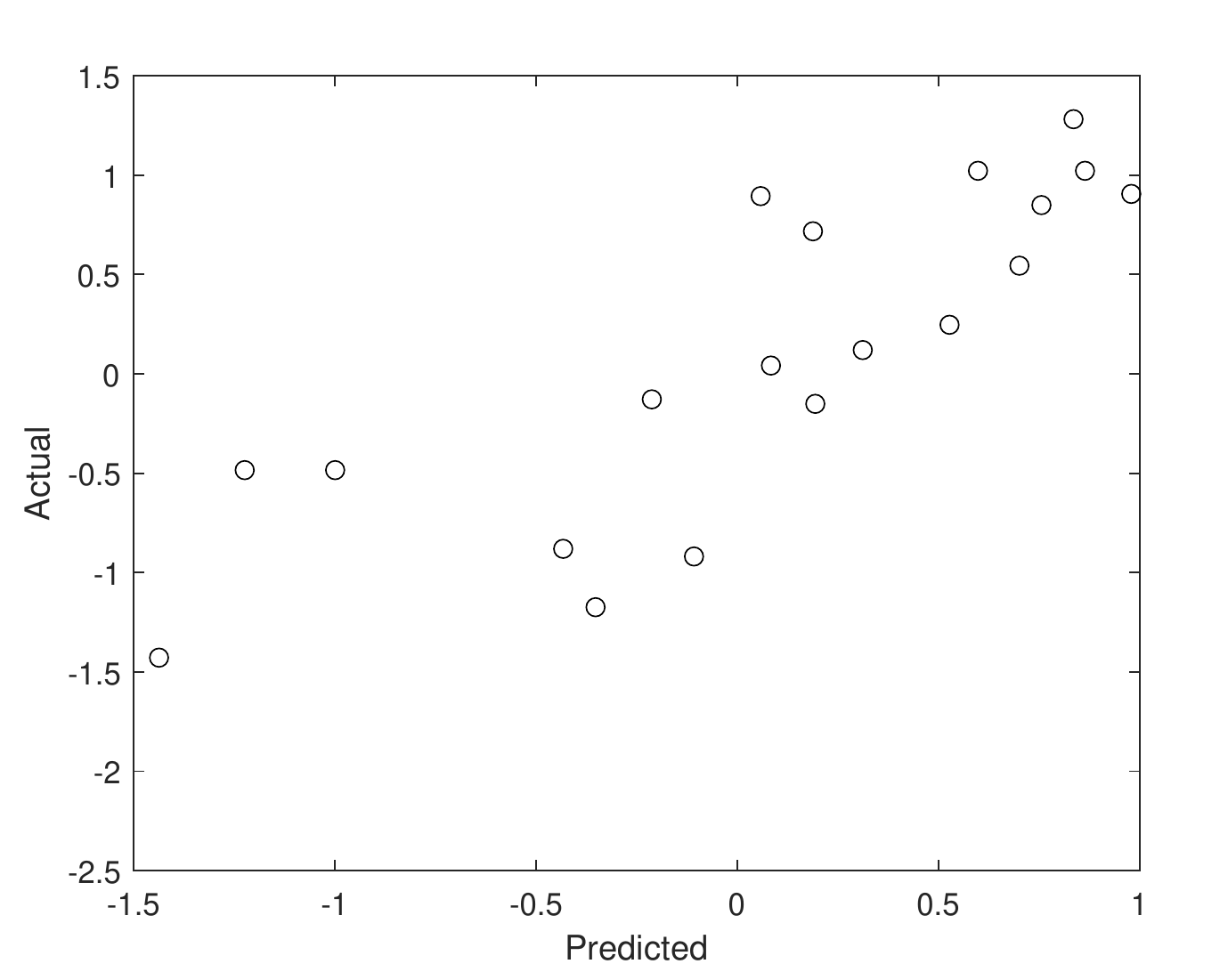}
	\end{minipage}%
}%
\subfigure[]{
	\begin{minipage}[t]{0.5\linewidth}
		\centering
		\includegraphics[width=\textwidth]{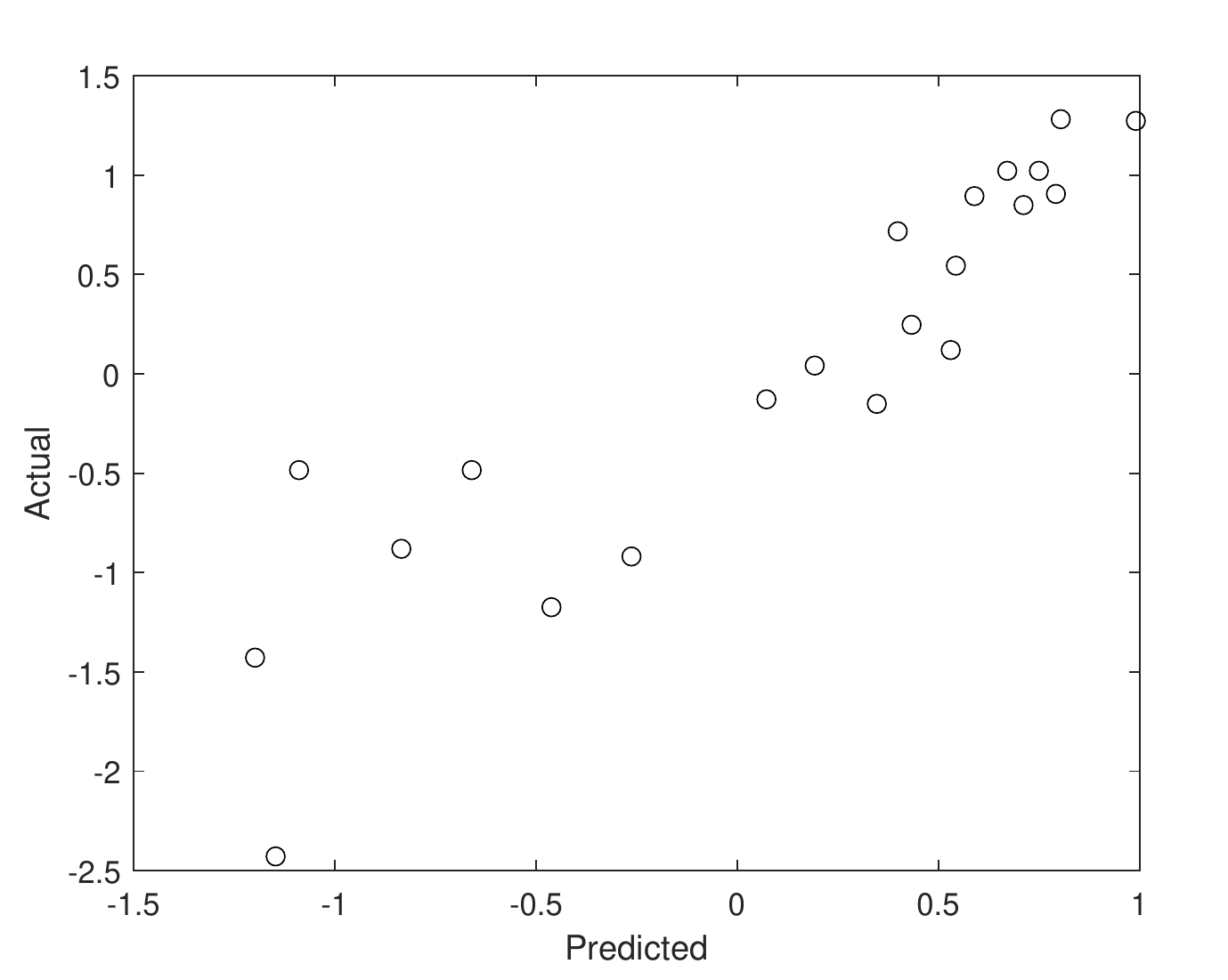}
	\end{minipage}%
}%
\caption{Panels (a) and (b) are the sufficient summary plots of \cite{tan2018convex} and our proposed method in the training set, respectively; Panels (c) and (d) are the scatterplots of \cite{tan2018convex} and our proposed method with the actual and predicted values for the testing samples, respectively.}
\label{Fig1}
\end{figure}

Figures \ref{Fig1}(a) and \ref{Fig1}(b) both show a good fit for both \cite{tan2018convex} and our proposed method in the training set data. Specifically, the method of \cite{tan2018convex} selects 23 genes with the adjusted $R^2$ 78.7\% while our proposed method only selects 21 genes with the adjusted $R^2$ 76.7\%. However, the predicted RMSE of \cite{tan2018convex} and our proposed method in the test set data are 2.192 and 2.068, respectively. The scatterplots of these two methods about the actual and predicted values for the 21 test samples are displayed in Figures \ref{Fig1}(c) and \ref{Fig1}(d). Thus in terms of prediction, our method is slightly better than \cite{tan2018convex}.

\section{Conclusion}
\label{sec:conc}
In this article, we extend the HSIC based SDR method of \cite{zhang2015direction} to handle a large $p$ and small $n$ scenario by borrowing the idea from \cite{tan2018convex}. The proposed method estimates the basis of the central subspace and performs sufficient variable selection simultaneously. Compared with other high-dimensional sparse SDR methods, our proposed method requires the weakest conditions so far. It enjoys a model free property and requires very mild conditions on $\X$ and no particular assumption on $Y|\X$, or $\X | Y$. The simulation studies showed that our method is highly efficient and stable in both $n>p$ and $n<p$ scenarios.

There are several possible prospects for future research. It may be of interest to extend this idea to multiple-index models, which is not trivial since it needs a new algorithm design. Moreover, the current computational bottleneck for our proposed method is on solving the majorization step, which  has a computational complexity of $O(p^3)$ per iteration. Thus, it will be also interesting to redesign a highly efficient algorithm such that our proposed method is scalable to accommodate large-scale data. Finally, the asymptotic properties for our method are deserved to discuss in the future which are not covered in this article.

\newpage

\appendix
%

\section{Some technical derivations}

\subsection{Proof of Proposition $\ref{prop:3.1.}$:}
We first compute the gradient function $\nabla f(\bPi)$.
Recalling the definition of $f(\bPi)$, we directly have
\begin{equation}\notag
\nabla f(\bPi)=\frac{1}{2n^2} \sum_{i,j=1}^{n} \exp\left(-\frac{ \langle \bPi, \Z_{ij} \rangle    }{2} \right)\tilde{L}_{ij}\Z_{ij}.
\end{equation}
Let us define a matrix ${\bf C}\in\bbR^{n\times n}$ with $\displaystyle c_{ij}=\exp\left(-\frac{ \langle \bPi, \Z_{ij} \rangle    }{2} \right)\tilde{L}_{ij}$ and $\X=[\X_1,\ldots,\X_n]^{\top}$, we have
\begin{eqnarray*}
\nabla f(\bPi) &=& \frac{1}{2n^2} \sum_{i,j=1}^{n} c_{ij}\Z_{ij}\\
              &=& \frac{1}{2n^2} \sum_{i,j=1}^{n} c_{ij}(\X_i-\X_j)(\X_i-\X_j)^{\top} \\
              &=& \frac{1}{2n^2} \sum_{i,j=1}^{n} c_{ij}\left(  \X_i\X_i^{\top} + \X_j\X_j^{\top}-\X_i\X_j^{\top}-\X_j\X_i^{\top}  \right) \\
              &=& \frac{1}{n^2}  \sum_{i,j=1}^{n} c_{ij}\left( \X_i\X_i^{\top} -\X_i\X_j^{\top}   \right)\\
              &=& \frac{1}{n^2} \X^{\top}\left(  \mbox{diag}({\bf C}{\bm 1}_{n})-  {\bf C}\right)\X,
\end{eqnarray*}
which establishes the first part of Proposition $\ref{prop:3.1.}$. Next, we prove the Lipschitz continuity of $\nabla f(\bPi)$ over the bounded set $\cD=\left\{ \bPi\in \cM,  {\rm tr}(\hat{\bSigma}^{1/2} \bPi\hat{\bSigma}^{1/2} )\leq 1 \right\}$. For any $\bPi\in\cD$ and $\tilde{\bPi}\in\cD$, by the triangle inequality, we obtain
\begin{eqnarray*}
\|\nabla f(\bPi)-\nabla f(\tilde{\bPi}) \|_{\rm F} &=& \| \frac{1}{2n^2} \sum_{i,j=1}^{n} \exp\left(-\frac{ \langle \bPi, \Z_{ij} \rangle    }{2} \right)\tilde{L}_{ij}\Z_{ij}-\frac{1}{2n^2} \sum_{i,j=1}^{n} \exp\left(-\frac{ \langle \tilde{\bPi}, \Z_{ij} \rangle    }{2} \right)\tilde{L}_{ij}\Z_{ij}  \|_{\rm F} \\
&\leq& \frac{1}{2n^2} \sum_{i,j=1}^{n} |\tilde{L}_{ij}|\| \Z_{ij}\|_{\rm F} \Big|\exp\left(-\frac{ \langle \bPi, \Z_{ij} \rangle    }{2} \right)- \exp\left(-\frac{ \langle \tilde{\bPi}, \Z_{ij} \rangle    }{2} \right)  \Big|\\
&\leq& \frac{1}{2n^2} \sum_{i,j=1}^{n} |\tilde{L}_{ij}|\| \Z_{ij}\|_{\rm F} \Big| \frac{ \langle \bPi-\tilde{\bPi}, \Z_{ij} \rangle    }{2}   \Big|,
\end{eqnarray*}
where the last inequality holds since $|e^x-e^y|\leq|x-y|$, for any $y\leq x\leq 0$. Further by the Cauchy-Schwartz inequality, we know $| \langle \bPi-\tilde{\bPi}, \Z_{ij} \rangle  |\leq \|\bPi-\tilde{\bPi}\|_{\rm F}\|\Z_{ij}\|_{\rm F}$. Thus, we finally get
\begin{eqnarray*}
\|\nabla f(\bPi)-\nabla f(\tilde{\bPi}) \|_{\rm F} &\leq& 	\frac{1}{4n^2} \sum_{i,j=1}^{n} |\tilde{L}_{ij}|\| \Z_{ij}\|_{\rm F}^2  \|\bPi-\tilde{\bPi}\|_{\rm F}\\
&=& \frac{\sum_{i,j=1}^{n} |\tilde{L}_{ij}|\| \Z_{ij}\|_{\rm F}^2}{4n^2} \|\bPi-\tilde{\bPi}\|_{\rm F},
\end{eqnarray*}
where $\displaystyle \frac{\sum_{i,j=1}^{n} |\tilde{L}_{ij}|\| \Z_{ij}\|_{\rm F}^2}{4n^2} $ is constant which verifies the claim.

\subsection{Linearized Alternating Direction Method of Multipliers Algorithm for Solving $(\ref{eqn3.6})$}
To implement the linearized alternating direction method of multipliers algorithm, we rewrite the subproblem in formula $(\ref{eqn3.6})$ as
\begin{equation}\notag
\begin{split}
&\underset{\bPi, \H\in \cM}{\min}\; \frac L2\Big\|\bPi-\left[ {\bf\Pi}^{(r)}-\frac 1L \nabla f({\bf\Pi}^{(r)}) \right]  \Big\|_{\rm F}^2 +\lambda \|\bPi\|_1+\infty\cdot \bI_{({\rm tr}(\H)>1)}, \\
&\mbox{   s.t. }\quad \hat{\bSigma}^{1/2} \bPi\hat{\bSigma}^{1/2} = \H.
\end{split}
\end{equation}
This is also equivalent to minimize the following scaled augmented Lagrangian function,
\begin{equation}\notag
\begin{split}
\cL_{\rho}(\bPi,\H,\bGamma)=&\frac L2\Big\|\bPi-\left[ {\bf\Pi}^{(r)}-\frac 1L \nabla f({\bf\Pi}^{(r)}) \right]  \Big\|_{\rm F}^2 + \lambda \|\bPi\|_1+\infty\cdot \bI_{({\rm tr}(\H)>1)}\\
&+\frac{\rho}{2}\| \hat{\bSigma}^{1/2} \bPi\hat{\bSigma}^{1/2} - \H+\bGamma  \|_{\rm F}^2,
\end{split}
\end{equation}
where $\rho$ is a small constant and $\bGamma$ is the dual variable. The L-ADMM minimizes the augmented Lagrangian function by alternatively solving one block of variables at a time. In particular, to update $\bPi$ at the $j$-th iteration, we need to minimize
\begin{equation}\notag
\frac L2\Big\|\bPi-\left[ {\bf\Pi}^{(r)}-\frac 1L \nabla f({\bf\Pi}^{(r)}) \right]  \Big\|_{\rm F}^2 +\lambda \|\bPi\|_1+\frac{\rho}{2}\| \hat{\bSigma}^{1/2} \bPi\hat{\bSigma}^{1/2} - \H_j+\bGamma_j  \|_{\rm F}^2,
\end{equation}
where $\H_j$ and $\bGamma_j$ are the $j$-th estimates of $\H$ and $\bGamma$ respectively. However, there is no closed-form solution for the above minimization problem. To tackle the difficulty, \cite{fang2015LADMM} proposed to linearize the quadratic term in the above problem by applying a second-order Taylor Expansion. Following the work of them, we obtain the update for $\bPi$:
\begin{equation}\notag
\begin{split}
\bPi_{j+1}=\underset{\bPi\in\cM}{\arg\min}\; & \frac L2\Big\|\bPi-\left[ {\bf\Pi}^{(r)}-\frac 1L \nabla f({\bf\Pi}^{(r)}) \right] \Big\|_{\rm F}^2 +\lambda \|\bPi\|_1 \\
&+\rho \langle \bPi-\bPi_j, \hat{\bSigma}\bPi_j \hat{\bSigma}-\hat{\bSigma}^{1/2} (\H_j-\bGamma_j)\hat{\bSigma}^{1/2} \rangle
+\frac \tau 2 \|\bPi-\bPi_j\|_{\rm F}^2.
\end{split}
\end{equation}
As suggested by \cite{fang2015LADMM}, we pick $\tau \geq 4\rho \lambda_{\rm max}^2(\hat{\bSigma})$ to ensure the convergence of the linearized alternating direction method of multipliers algorithm. The above iterate can be written in the more familiar notation:
\begin{equation}\notag
\begin{split}
\bPi_{j+1}=\underset{\bPi\in\cM}{\arg\min}\; & \frac{L+\tau}{2}  \Big\|\bPi-\Big( \frac{\tau}{L+\tau} \left[ \bPi_{j}-\frac{\rho}{\tau}\hat{\bSigma}\bPi_{j}\hat{\bSigma}+\frac{\rho}{\tau}\hat{\bSigma}^{1/2}(\H_j-\bGamma_{j})\hat{\bSigma}^{1/2} \right] \\
&+\frac{L}{L+\tau} \left[ \bPi^{(r)}-\frac{\nabla f(\bPi^{(r)})}{L}\right] \Big)  \Big\|_{\rm F}^2 +\lambda \|\bPi\|_1
\end{split}	
\end{equation}
which has the closed-form solution
\begin{equation*}
\bPi_{j+1}	= {\rm Soft}\left( \frac{\tau}{L+\tau} \left[ \bPi_{j}-\frac{\rho}{\tau}\hat{\bSigma}\bPi_{j}\hat{\bSigma}+\frac{\rho}{\tau}\hat{\bSigma}^{1/2}(\H_j-\bGamma_{j})\hat{\bSigma}^{1/2} \right] +\frac{L}{L+\tau} \left[ \bPi^{(r)}-\frac{\nabla f(\bPi^{(r)})}{L}\right],\frac{\lambda}{L+\tau}
 \right) ,
\end{equation*}
where Soft is the element-wise soft-thresholding to a matrix: ${\rm Soft}(A_{ij},b)={\rm sign}(A_{ij})\max(|A_{ij}|-b,0)$. Next, the update of $\H$ can be obtained as
\begin{equation*}
\H_{j+1} =\underset{ \H\in\cM, {\rm tr}(\H)\leq1 }{\arg\min }\; \frac{1}{2}\| \H-(\hat{\bSigma}^{1/2}\bPi_{j+1}\hat{\bSigma}^{1/2}+\bGamma_{j}) \|_{\rm F}^2,
\end{equation*}
which has a closed-form solution according to the following proposition.
\begin{proposition}
\label{proA.1}
Let $\cF=\left\{ \H\in\cM: {\rm tr}(\H)\leq1  \right\}$ and $\displaystyle P_{\cF}({\bf W})=\underset{ \H\in\cF }{\arg\min }\; \frac{1}{2}\|\H-{\bf W}\|_{\rm F}^2.$ If ${\bf W}=\sum_{i=1}^{p}\omega_{i}\bf{u}_{i}\bf{u}_{i}^{\top}$ is a spectral decomposition of $\bf{W}$, then $P_{\cF}({\bf W})=\sum_{i=1}^{p}(\omega_{i}-\theta^{*})_{+}u_{i}u_{i}^{\top}$, where $(\omega_{i}-\theta)_{+}=\max(\omega_i-\theta,0)$ and $\theta^{*}$ is the minimum value satisfying $\sum_{i=1}^{p}(\omega_{i}-\theta)_{+}\leq 1.$
\end{proposition}
The above proposition follows directly from Lemma 4.1 in \cite{vu2013fantope}, Proposition 10.2 in \cite{gao2017sparse}, and Proposition 1 in the Appendix of \cite{tan2018convex}. Thus,  by Proposition \ref{proA.1}, we have
$$\H_{j+1}=P_{\cF}(\hat{\bSigma}^{1/2}\bPi_{j+1}\hat{\bSigma}^{1/2}+\bGamma_{j}).$$
Finally, we update the dual variable by
$$\bGamma_{j+1} = \bGamma_{j}+ \hat{\bSigma}^{1/2}\bPi_{j+1}\hat{\bSigma}^{1/2}-\H_{j+1}. $$

\bigskip

\begin{center}
	{\large\bf ACKNOWLEDGEMENTS}
\end{center}
Chen's research was supported by SUSTech startup funding.

\bibliographystyle{JASA}
\bibliography{Bibliography-MM-MC}

\end{document}